\documentclass{article}
\usepackage{spconf,amsmath,graphicx}
\usepackage{amssymb}
\usepackage{placeins}
\usepackage{latexsym}
\usepackage{enumitem}
\setlist{nosep, leftmargin=14pt}

\usepackage{mwe} 

\title{Paired Diffusion: Generation of related, synthetic PET-CT-Segmentation scans using Linked Denoising Diffusion Probabilistic Models}
\name{Rowan Bradbury$^{\dag1}$, Katherine A. Vallis$^{\dag^1}$, Bart\l omiej W. Papie\.z$^{\dag^2}$}

\address{$^1$Department of Oncology, University of Oxford, Oxford\\$^2$Big Data Institute, Li Ka Shing Centre for Health Information and Discovery, University of Oxford, Oxford}

\begin{document}
\maketitle
\begin{abstract}
The rapid advancement of Artificial Intelligence (AI) in biomedical imaging and radiotherapy is hindered by the limited availability of large imaging data repositories. With recent research and improvements in denoising diffusion probabilistic models (DDPM), high quality synthetic medical scans are now possible. Despite this, there is currently no way of generating multiple related images, such as a corresponding ground truth which can be used to train models, so synthetic scans are often manually annotated before use.
This research introduces a novel architecture that is able to generate multiple, related PET-CT-tumour mask pairs using paired networks and conditional encoders. Our approach includes innovative, time step-controlled mechanisms and a `noise-seeding' strategy to improve DDPM sampling consistency. While our model requires a modified perceptual loss function to ensure accurate feature alignment we show generation of clearly aligned synthetic images and improvement in segmentation accuracy with generated images.

\end{abstract}
\section{Introduction}
\label{sec:intro}
In radiotherapy planning, segmentation of tumours and organs-at-risk (OARs) is paramount for treatment efficacy and safety \cite{segaccoutcome}. Despite the sophistication of manual delineation, it remains fraught with inter-observer variability and is challenged by the necessity for extensive professional training and verification \cite{duke2021assessing, evalRTvoldelin}. Whilst deep learning has produced impressive results, the biggest bottleneck of these methods is commonly agreed to be the lack of training data available as a result of privacy concerns and the specificity required for different tumour types. Advances in recent years have often been attributed to an increase in the availability of annotated data that can train models \cite{Cardenas2019}. More recent architectures like transformers are the new state-of-the-art in other fields where millions of training samples are accessible, surpassing convolutional approaches by capturing global dependencies. Yet, this feature makes them susceptible to over-fitting in data-constrained scenarios such as medical imaging, inhibiting full training and utilisation \cite{octreehecktor}. This is exacerbated in tumour segmentation, where the intricacy and variability of tumour morphology demands a high volume of diverse training samples to achieve accurate generalisation.

DDPMs emerge as a solution with their exceptional fidelity in image generation, prompting their use in creating synthetic medical datasets to improve network generalisation and reduce dataset imbalance with freely shareable data \cite{DDPM}. Despite impressive demonstrated generation capability, existing DDPM applications in medical imaging have been confined to pretraining due to their inability to generate the associated ground truth segmentation for a generated scan, or additional modalities, such as corresponding Positron Emission Tomography (PET) for a generated Computed Tomography (CT) image that is required to train a supervised network \cite{medicaldiffusion}. Current workarounds involve style transfer DDPM networks to generate the additional images, such as generating scans from an existing segmentation - but these have various limitations, including being limited by the number of real input images and the inability to force consistency between more than two images \cite{segtoimg}.

This study aimed to address this gap and synthesise related `paired' medical images and annotations that can be used to directly train downstream AIs and reduce risk of overtraining. Our contributions are as follows: first, a novel multi-model architecture was built using a multimodal Fludeoxyglucose (FDG)-PET-CT/Tumour segmentation dataset \cite{hecktorbook}. By allowing cross-modal guidance during the diffusion process, we not only facilitate the generation of diverse and realistic datasets but also ensure the alignment of each modality to the other without harming generation fidelity. This is of the first studies to explore generation of entirely-synthetic annotated data in medical imaging,  and one which supports more than two linked modalities \cite{braintumoursegganvddpm,pham,segtoimg}.
\section{Methodology}
\label{sec:majhead}
\subsection{Denoising Diffusion Probabilistic Models}
DDPMs are generative models that create images iteratively from random noise \cite{DDPM}. Training involves applying noise with a forward diffusion process and then learning to reverse various levels of noise `damage' to training images, this noising/denoising is done in steps from fully denoised at time 0 ($T_0$) to fully noised after $T$ steps (typically 1000). During training the model takes image $x$ and noises it according to a random time step $T_t$ with a known noise forming noised image $x_t$. The model predicts the noise added to $x$, which is compared to the known noise. In sampling this predicted noise is typically scaled and subtracted to approximate the image denoised one step ($x_{t-1}$). When sampling, a model can take an entirely random noise (equivalent to fully noised image $x_T$) and forms a new image by iterating this step-wise process to $T_0$.

\subsection{Data and Data Preprocessing}
\label{ssec:subhead}
We utilised two datasets: the Head and Neck Segmentation (HaN-Seg) challenge dataset for OAR segmentation and unpaired generation testing, containing 56 3D CT/Magnetic Resonance (MR) image pairs with OAR labels \cite{hanseg}, and the MICCAI Head and Neck Tumour Segmentation (HECKTOR) 2022 dataset, featuring 524 cases with FDG-PET-CT pairs and annotations \cite{hecktorbook}.

Images were resampled to a uniform 2x2x2 mm$^3$ voxel size with trilinear interpolation for CT, and PET \cite{Chu2023,bartekhecktor}. Segmentations were processed with nearest neighbour interpolation to preserve their binary nature. To accommodate the varying resolutions of data sourced from multiple centres, we aligned images using the encoded origin of each array, applying padding vertically where necessary and cropping to maintain a uniform slice. 2D slices were taken wherever tumour is visible; we employed a \textit{smart} crop function taking 128x128 pixel slices, prioritising the `centre of mass' of non-background voxels over geometric centre. To mitigate over-fitting, the training pipeline included data augmentation techniques such as random jitter, rotation, and a random flip across the mid-sagittal plane. The datasets were partitioned into training, validation, and test sets in an 80:10:10 ratio.

\subsection{Model Design}
\label{sec:print}
Our model follows a multi-network structure, each with its own optimiser, but shared losses. The multi-network structure enables full specialisation for the individual modality type and provides hugely improved quality over a single DDPM with multi-channel outputs.
Drawing inspiration from MedSegDiff V1 \cite{medsegdiff}, each sub-network employs $n$ conditional encoders to guide the diffusion process according to the other $n$ modalities, with simple attention mechanisms driving aligned feature extraction between both encoders. Latent features are directly concatenated and we introduce a timestep-aware convolutional block attention module (CBAM) which can dynamically shift the impact of the conditional layers at each timestep. This aspect may be particularly effective in emphasising large-scale features at early timesteps and refining finer details as the model progresses which, in-line with the noise schedule, is how diffusion models generate images.

 \begin{figure}[htb]
\begin{minipage}[b]{1.0\linewidth}
  \centering
  \centerline{\includegraphics[width=8.5cm]{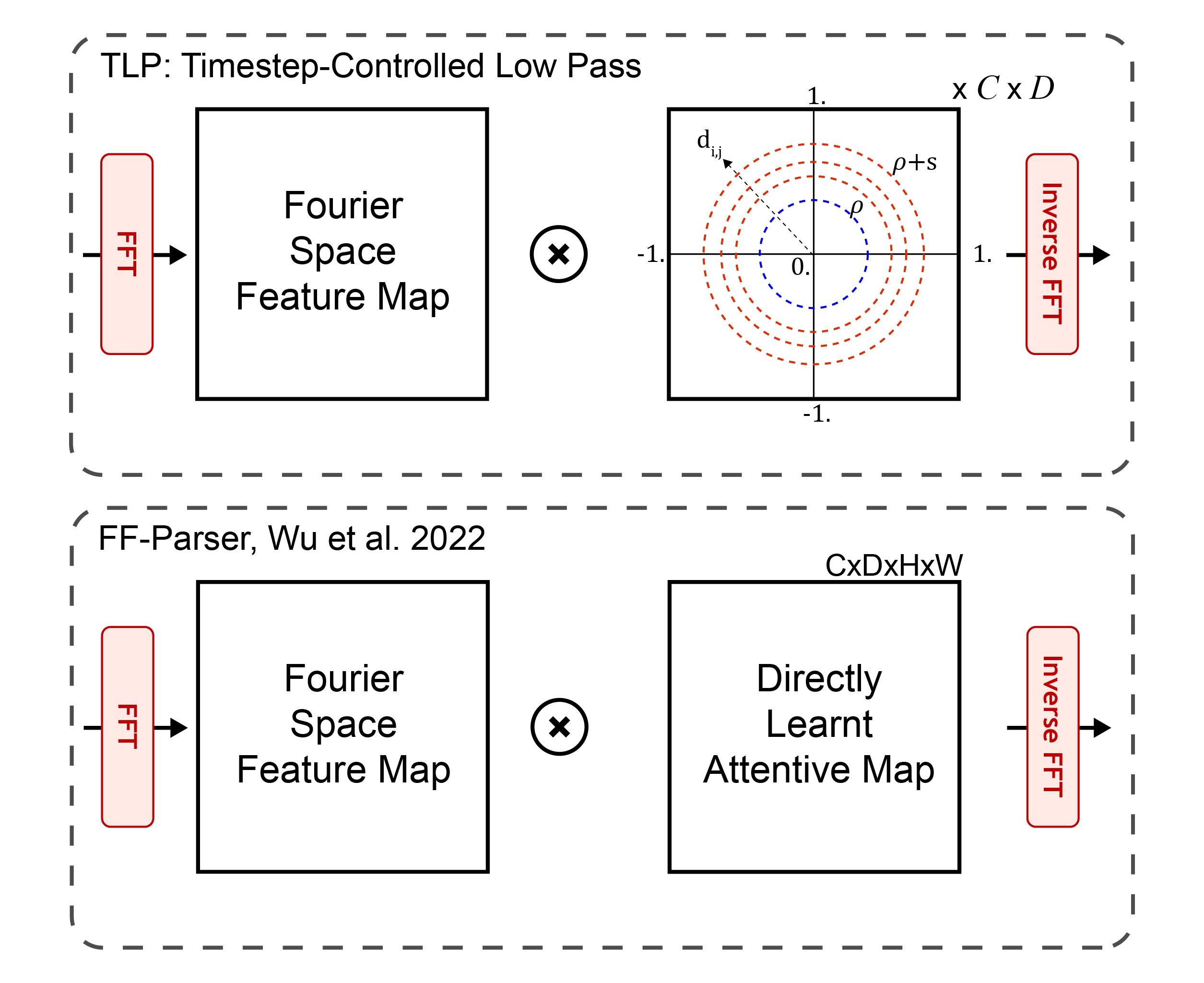}}
  \centerline{(a) FF-Parser vs Timestep-Controlled Low Pass}\medskip
\end{minipage}
\caption{Illustration of the simplified low-pass filter, parameterised by just two variables ranging from 0 to 1. This simpler approach makes it more intuitive to control and faster to learn useful representations compared to learning a full-width, height, and channel size filter. In our model, this filter plays a vital role in diminishing noise from the conditional input (as it is also noised) to focus on semantic features.
As all information into the conditional encoders has to pass through this layer we found the non-pretrained FF-Parser to introduce training difficulties by degrading the input, in comparison to our method which enforces noise reduction from initialisation with only two learnable weights. With the Gaussian noise being added across all frequencies we did not believe the ability of FF-Parser to generate more complex noise-reduction schemes useful here.
Crucially our filter also allows for timestep-dependent adjustments, enhancing its denoising capability significantly to effectively and efficiently reduce input noise regardless of timestep. 
}
\label{ffp}
\end{figure}
MedSegDiff utilises an attentive map multiplied in frequency space that learns a low pass blurring filter. The rationale here is to remove noise we are trying to predict from $x_t$ being transferred into the conditional encoder so it can focus on semantic features. Our guiding condition is also a noised image from other model outputs, so we place FF-Parser-like blocks to denoise them before entry to the model.
We innovated the FFParser by parameterising the low pass filter into an input-size agnostic low memory filter, which forces immediate functionality and quick learning in such a critical position. Furthermore, this enables control from a timestep-aware-MLP to dynamically reduce noise depending on the level of noise that has been added. Sampled results from this mechanism show ideal noise reduction at all timesteps, compared to the static and difficult to learn FF-Parser which resulted in overblurring and loss of semantic information at low timesteps, and underblurring at high steps.

\subsection{Forward Diffusion Modification and Perceptual Loss}
\label{ssec:subhead}
We observe a distinct bias, making learning dark image significantly slower, especially at low resolutions. The difficulty in forcing generation of largely bright or dark images has been noted by consumer users of open-source diffusion software like stable diffusion. We note that uniform addition of a random constant to the forward diffusion noise dynamically changes the brightness of the image and forces the model to learn the inherent `brightness' of the training images, greatly enhancing ability to learn dark images like medical images \cite{offsetnoise}.
Standard noise is sampled per pixel, width and height coordinates $i,j$, from a standard normal distribution such that $
N_{i,j} \sim \mathcal{N}(0, 1)$. A constant value, $C$, is also sampled from the standard normal distribution, scaled, and uniformly added to each pixel in matrix $N$. This adjusted noise matrix $N'$ can then be used as standard in forward diffusion. Scaling with weights larger than 0.1 were observed to harm sample quality.
\begin{equation}
C \sim 0.1 \cdot \mathcal{N}(0, 1)
\end{equation}
\begin{equation}
N'_{i,j} = N_{i,j} + C \quad \forall i,j
\end{equation}

To drive importance of conditional features we implement a rudimentary perceptual loss function with pre-trained VGG-16 based on the outputs on all models.
If added too early this inhibits learning as it encourages all images to be identical, so this was implemented past epoch 50 only. Additionally we implement the inverse of the forward diffusion function, which approximates the fully denoised image $x_0$ from model predictions, providing a significantly cleaner input to perceptual loss functions.

\subsection{Noise Seeding and Testing Synthetic Data}
\begin{figure}[htb]
\begin{minipage}[b]{0.48\linewidth}
  \centering
  \includegraphics[width=0.75\linewidth]{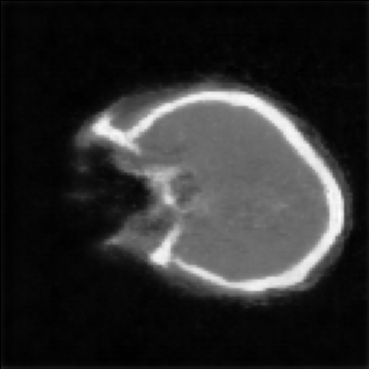}
 
  \label{fig:real}
\end{minipage}
\hfill
\begin{minipage}[b]{0.48\linewidth}
  \centering
  \includegraphics[width=0.75\linewidth]{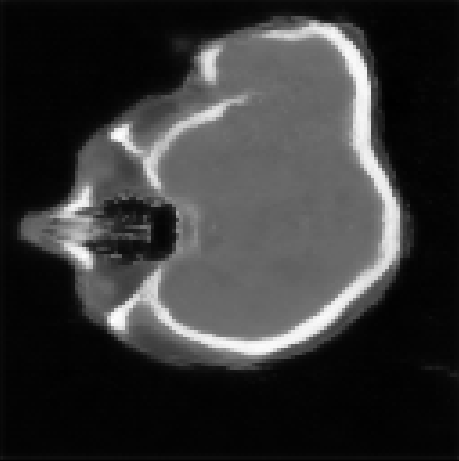}
  \label{fig:synthetic}
\end{minipage}
\caption{Examples of misgenerations caused by `fast' sampling too aggressively.}
\label{fig:misgen}
\end{figure}
DDPM excels in generating high-quality images but is limited by slow generation speeds. Fast sampling techniques, such as DPM-solver++ aim to attenuate this issue, but overly aggressive sampling can lead to misgenerations including abnormal skull shapes \cite{dpmsolver}. Our observations suggest that subtle alterations in the input noise—such as imperceptible darkening or introducing a regular pattern akin to the target shape—can influence the generation outcome, triggering misgenerations or promoting successful ones. We hypothesise that the initial phase of DDPM generation, involving the creation of low-frequency, high-level structures is the most challenging. This is evident in the initial poor $x_0$ predictions that very rapidly improve with progression. We hypothesise that \textit{good} and \textit{bad} seeds exist depending on if the underlying patterns in the noise align with the generation goals.

In order to force better generations and reduce encountering bad random starting points we noise a real image to 80-90\% of the noise schedule, human-indistinguishable from random noise. Applying the diffusion process from here we effectively `seed' the image. This technique resembles a very mild image-to-image diffusion process. The resultant images, while distinct from their source, often retain high-level details like the specific slice location, allowing for more aggressive sampling methods without the risk of misgenerations. 
In our experiments we assess if seeding with real images from 80\% of the noise schedule forces alignment of the resulting images.  To test the viability of the generated data, an OAR autocontouring network (TransUNet) was pretrained with the original data, synthetic data only or not at all and then trained for pituitary gland segmentation. Similarly we also trained a PET-CT and CT only network with entirely synthetic pairs, real pairs or synthetic pairs seeded from the real pairs.

\section{Results}
 \begin{figure}[!htb]

\begin{minipage}[b]{1.0\linewidth}
  \centering
  \centerline{\includegraphics[width=8.5cm]{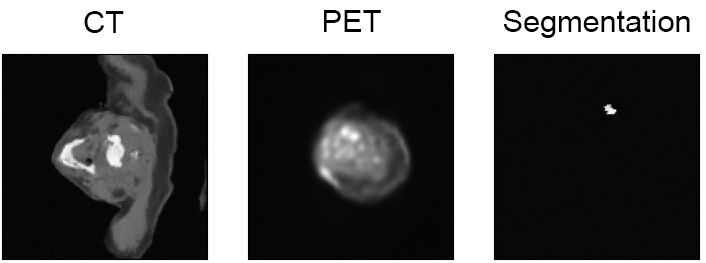}}
  \centerline{(a) Non-Aligned Generations (No Perceptual Loss)}\medskip
\end{minipage}
\begin{minipage}[b]{1.0\linewidth}
  \centering
  \centerline{\includegraphics[width=8.5cm]{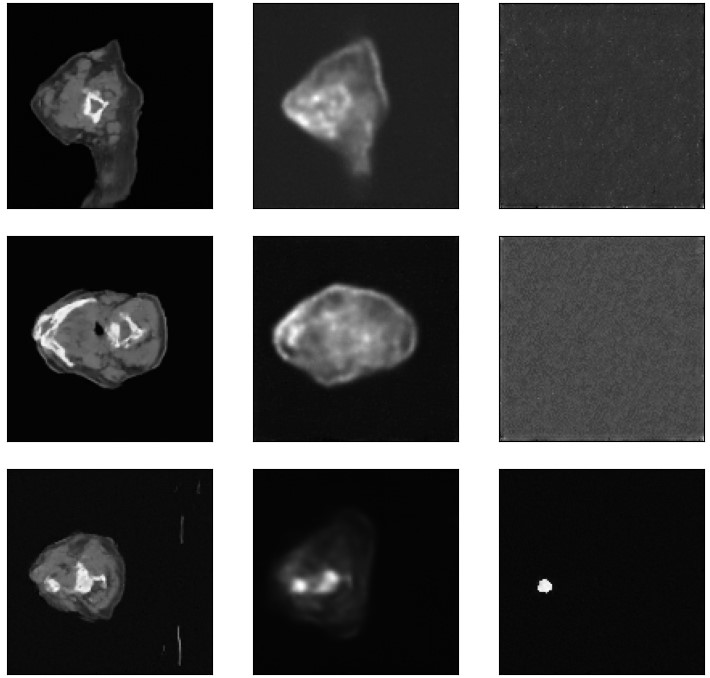}}
  \centerline{(b) Aligned Generations (Perceptual Loss)}\medskip
\end{minipage}
\begin{minipage}[b]{1.0\linewidth}
  \centering
  \centerline{\includegraphics[width=8.5cm]{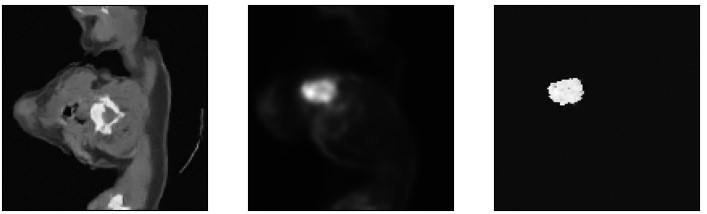}}
  \centerline{(c) Aligned Generations (Seeded Perceptual Loss)}\medskip
\end{minipage}
\caption{Examples of image generation with different model training configurations. Columns represent each synthetic modality. Left to right: CT, PET, tumour segmentation.}
\label{modelresults}
\end{figure}

\begin{figure}[htb]

\begin{minipage}[b]{0.48\linewidth}
  \centering
  \includegraphics[width=0.75\linewidth]{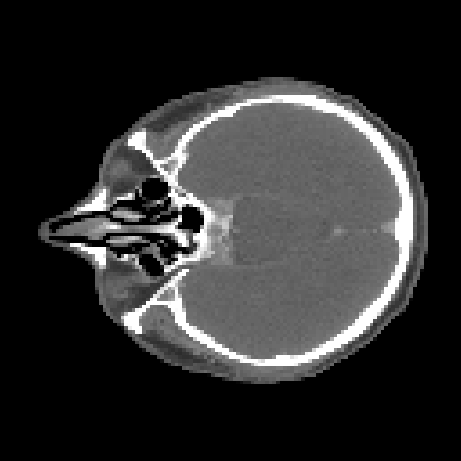}
  \centerline{(a) Real CT scan}
  \label{fig:real}
\end{minipage}
\hfill
\begin{minipage}[b]{0.48\linewidth}
  \centering
  \includegraphics[width=0.75\linewidth]{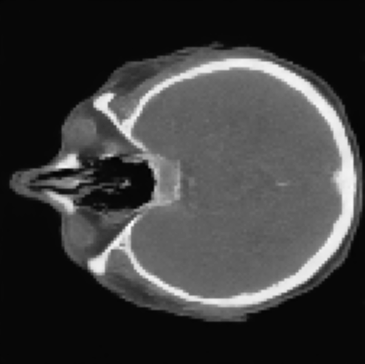}
  \centerline{(b) Synthetic Image}
  \label{fig:synthetic}
\end{minipage}

\caption[Real vs Synthetic CT scan]{Comparison of fully-trained, single-modality DDPM generated images to assess generation quality.}
\label{fig:realorfake}
\end{figure}

Without perceptual loss or with conditional feature pathways severed, the generation of images with the model is not aligned at all, even in general slice positioning (Fig. \ref{modelresults}a). This setup has shared loss and is akin to Pham et al. who claim paired results \cite{pham}. Noise seeding here does not enforce any paired generation either.
With perceptual loss, images are encouraged to have similar features (Fig. \ref{modelresults}b). This encourages `pairedness' through the conditional encoder, but also leads to decreased image quality (especially in segmentation channel) and alignment of literal features that do not make semantic sense across modalities. For example the alignment of bright spots across PET-CT encourages tumour to form where bright bone shows on CT and inversely bright 'bone' to show where the tumour is on PET. Whilst the perceptual loss seems to especially harm the segmentation generation, it is robust when the tumour region has high contrast on the PET scan.
Using noise seeding in conjunction with the perceptual loss restores image quality and enforces a good overall structure whilst being different to the input image (Fig. \ref{modelresults}c).
Note that due to compute limitations these models were not fully trained, the single-modality model is significantly smaller and faster to train, a sample of the generation quality is shown in Fig.~\ref{fig:realorfake}.

\subsection{Use of Synthetic Data}
Synthetic data massively increased the speed of convergence in OAR contouring and also led to a lower minimum validation loss. Segmentation accuracy was tested with the trained models on an unseen test set and a one-way ANOVA was carried out, finding a statistically significant difference between the groups (F value 4.99, p=0.0037 $<$ 0.05). As post-hoc, t-tests were carried out with Bonferroni correction for multiple comparisons, finding significance only between the control and synthetic-pretrained networks (t value -3.93, p=0.00046 $<$ 0.05). No significant difference was found with directly training a whole-slice PET-CT tumour segmentation network with the paired data, but mild improvement was seen in the Dice score with the harder CT only task that was prone to overtraining on the real data (t-test, t value -2.57, p = 0.01 $<$ 0.05).

\section{Discussion}
Our results show that our model is capable of generating realistic paired images with the help of noise seeding. However, the simple perceptual loss function alone leads to a decrease in image quality and causes unrealistic images because of the alignment of unrelated features. Whilst we have attempted to train a ViT-based Siamese network with modality-specific heads to perform a more `smart' approach to this, the network does not react well to poor $x_0$ predictions - having been trained on clean data only. Generating similar, aligned data to train it on beforehand is not possible, especially for earlier timesteps, which are the most important.

A few other works have arisen recently on this topic, notably Dorjsembe et al  show a network that generates MR images based on pre-existing segmentations \cite{segtoimg}. This is not purely synthetic and is limited to using existing segmentations, which can be positionally translated for variation. Unlike our design however, modalities are unable to \textit{push-back} on unrealistic positioning. Further, there is no way to generate two images like PET and CT from one segmentation and ensure they are mutually aligned.

To further show the power of this synthetic data supplementation on segmentation, a model should be trained on a tumour bounding box rather than a whole slice and in a difficult scenario that promotes overtraining such as mask transfiner \cite{octreehecktor}.  Future work should also assess our informal discoveries regarding noise seeding with quantitative metrics to assess the reduction in variance this may cause. For entirely synthetic data, faster sampling network designs like General Adversarial Networks or Variational Auto Encoders can be overtrained and used to seed these networks whilst having high frequency information rebuilt from scratch with DDPM-like quality.

\FloatBarrier

\section{Compliance with ethical standards}
\label{sec:ethics}
This work is a study for which no ethical approval was required.

\section{Acknowledgments}
\label{sec:acknowledgments}
KV acknowledges support from CRUK C6078/A28736.

\bibliographystyle{IEEEbib}
\bibliography{refs}

\end{document}